\documentclass{ifacconf}

\usepackage{amsfonts}

\usepackage{graphicx}      
\def\FigPath{Figures/}
\usepackage{multirow,array,graphicx}
\usepackage{tabularx,booktabs}
\usepackage[table]{xcolor}

\definecolor{TabShade}{rgb}{0.00000,0.38039,0.39608}
\newcommand{\Hi}{\cellcolor{TabShade!20}}

\newcommand{\reftab}[1]{Table~\ref{#1}}
\newcommand{\refeq}[1]{(\ref{#1})}
\newcommand{\reffig}[1]{Fig.~\ref{#1}}

\newcommand{\refsec}[1]{Section~\ref{#1}}

\usepackage{physics}
\newcommand{\matr}[1]{\boldsymbol{#1}}
\newcommand{\vek}[1]{\boldsymbol{#1}}

\newcommand{\vtx}[2]{{#1}_{\rm{#2}}} 

\DeclareSymbolFont{matha}{OML}{txmi}{m}{it}

\newcommand{\pdrev}[2]{\dfrac{\partial #1}{\partial #2}}
\newcommand{\drev}[2]{\dfrac{d #1}{d #2}}
\newcommand{\Rey}{\operatorname{\mathit{R\kern-.04em e}} }
\newcommand{\Pecl}{\operatorname{\mathit{P\kern-.08em e}}}
\newcommand{\Pran}{\operatorname{\mathit{P\kern-.03em r}}}
\newcommand{\Rayl}{\operatorname{\mathit{R\kern-.04em a}}}
\newcommand{\Nuss}{\operatorname{\mathit{N\kern-.09em u}}}
\newcommand{\Gras}{\operatorname{\mathit{G\kern-.05em r}}}

\renewcommand{\norm}[1]{\lVert#1\rVert}

\DeclareMathSymbol{\varv}{\mathord}{matha}{118}
\newcommand{\cliq}{c}
\newcommand{\mHup}{m_{h}}

\newcommand{\mdair}{\vtx{\dot{m}}{a}}
\newcommand{\avib}{\vtx{a}{vib}}
\newcommand{\Tair}{\vtx{T}{a}}
\newcommand{\GPIn}{\vek{\vartheta}}
\newcommand{\fTor}{\zeta}
\newcommand{\vPDE}{v}
\newcommand{\DPDE}{D}

\newcommand{\msin}{\vtx{\dot{m}}{s}}
\newcommand{\mliq}{\vtx{\dot{m}}{l}}
\newcommand{\YDelta}{\Delta Y}

\newcommand{\Tsat}{\vtx{T}{s}}
\newcommand{\Tdew}{\vtx{\varphi}{a}}
\newcommand{\Pair}{\vtx{P}{a}}

\newcommand{\Svar}{z}

\newcommand{\ubilin}{{\vek{h}}}
\newcommand{\Uvec}{{\hat{\vek{u}}}}
\newcommand{\xobs}{\vek{\eta}}
\newcommand{\dxobs}{\dot{\vek{\eta}}}
\newcommand{\xobsr}{\hat{\vek{\eta}}}
\newcommand{\dxobsr}{\dot{\hat{\vek{\eta}}}}
\newcommand{\Pmat}{\matr{P}}

\newcommand{\Cobs}{\hat{\matr{C}}}

\newcommand{\udim}{5}

\newcommand{\np}{N}
\newcommand{\nr}{r}

\newcommand{\bigzero}{\mbox{\normalfont\large\bfseries 0}}

\newcommand{\Imat}[1]{ \matr{I}_{#1 \times #1} }

\newcommand{\xr}  {\hat{\vek{x}}_1}
\newcommand{\xdr} {\dot{\hat{\vek{x}}}_1}
\newcommand{\Ar}  { \vek{A}_r }
\newcommand{\Qr}  { \vek{\mathcal{Q}}^{[1]}_r }
\newcommand{\Qrf} { \vek{T} \vek{\mathcal{Q}}^{[1]} (\Imat{5}\otimes \vek{V}) }

\newcommand{\Br}  { \vek{B}_r }
\newcommand{\Cr}  { \vek{C}_r }

\newcommand{\Qf}{ \vek{\mathcal{Q}}^{[1]} }

\usepackage{siunitx}
\newcommand{\refappendix}[1]{\hyperref[#1]{Appendix~\ref*{#1}}}

\usepackage{pgf,tikz,pgfplots}
\pgfplotsset{compat=newest}
\usetikzlibrary{arrows,decorations.markings}
\usetikzlibrary{positioning}
\usepackage{tikz-network}
\usetikzlibrary{plotmarks}
\usepackage{algorithm}
\usepackage{algpseudocode}
\makeatletter
\let\old@ssect\@ssect 
\makeatother
\usepackage{natbib}
\usepackage{hyperref}
\makeatletter
\def\@ssect#1#2#3#4#5#6{%
  \NR@gettitle{#6}
  \old@ssect{#1}{#2}{#3}{#4}{#5}{#6}
}
\makeatother

\usepackage{pgf}
\definecolor{lime}{HTML}{A6CE39}

\definecolor{mycolor1}{rgb}{0.25098,0.49804,0.71765}%

\usepackage[customcolors]{hf-tikz}
\usetikzlibrary{fit}
\tikzset{style green/.style={
		set fill color=green!20!lime!60,
		set border color=white,
	},
	style cyan/.style={
		set fill color=cyan!90!blue!60,
		set border color=white,
	},
	style orange/.style={
		set fill color=orange!80!red!60,
		set border color=white,
	},
	hor/.style={
		above left offset={-0.15,0.31},
		below right offset={0.15,-0.125},
		#1
	},
	ver/.style={
		above left offset={-0.1,0.3},
		below right offset={0.15,-0.15},
		#1
	}
}

\tikzset{%
	highlight/.style={rectangle,rounded corners,fill=red!80,fill opacity=0.3,thin,inner sep=0pt}
}

\graphicspath{{Figures/}}

\newcommand\copyrighttext{%
  
  \textcopyright 2022 the authors. This work has been accepted to IFAC for publication under a Creative Commons Licence CC-BY-NC-ND.}
\newcommand\copyrightnotice{%
\begin{tikzpicture}[remember picture,overlay]
\node[anchor=north,yshift=-10pt] at (current page.north) {\shortstack{Accepted for publication at the IFAC Workshop on Control of Systems \\ Governed by Partial Differential Equations (CPDE) 2022\\ \\
		\fbox{\parbox{\dimexpr\textwidth-\fboxsep-\fboxrule\relax}{\copyrighttext}}} };
\end{tikzpicture}%
}

\begin{document}
\copyrightnotice

\begin{frontmatter}

\title{Reduction and Observer Design for a Grey-Box Model in Continuous Pharmaceutical Manufacturing\thanksref{footnoteinfo} }

 \thanks[footnoteinfo]{We would like to express our gratitude to Dr. Robin Meier and the R\&D team of L.B. Bohle Maschinen und Verfahren GmbH for providing the experimental data.}

\author[IRT]{Ahmed Elkhashap} 
\author[IRT]{Dirk Abel} 

\address[IRT]{Institute of Automatic Control, RWTH Aachen University, Aachen, Germany (e-mail: A.Elkhashap@irt.rwth-aachen.de)}

\begin{abstract}                
In this contribution, a novel Reduced Order Model (ROM) formulation of the 
grey-box model proposed in \cite{ELKHASHAP2020a} for the pharmaceutical continuous vibrated fluid bed dryer (VFBD) is presented. The ROM exploits the $\mathcal{H}_2$-norm projection-based model order reduction method after a special solution formulation of the model's infinite-dimensional part. This is mainly by introducing a vector field mapping between the model parts casting the semi- discretized PDE into a bilinear form. 
The ROM produced is then integrated into an nonlinear Kalman Filtering-based observer design also handling the estimation of the model's algebraic variables. Evaluations of the FOM, ROM, ROM-based observer variants, and the FOM-based observer are performed using Monte-Carlo simulations as well as simulations based on experimental data of the real system. It is shown that the ROM could reproduce the FOM states accurately with a relative mean square error below $0.3\,\%$ for the experimental data simulation. This is while reaching a computational-time reduction up to a factor of $40$. The ROM-based observer with algebraic states correction is shown (using Monte-Carlo simulations) to be able to converge to the true values for all cases regardless of initialization. Moreover, it is also shown that the performance degradation of the observer due to reduction is practically insignificant. This is while the computational speedup of the observer due to reduction reached a factor of more than third order of magnitude. 
\end{abstract}

\begin{keyword}
Continuous Pharmaceutical Manufacturing, Model Order Reduction, Observer Design, Early Lumping, Distributed Parameter Systems 
\end{keyword}

\end{frontmatter}

\section{Introduction}
Continuous Pharmaceutical Manufacturing (CPM) gained significant attention within the last decade. This manifested mainly into focused research efforts noticeable within a wide spectrum of engineering and natural sciences disciplines related to pharmaceutical manufacturing, \cite{Rantanen.2015}. The attention captured by CPM can be well justified. This is due to the enormous potential of the new dynamic manufacturing strategy regarding the manufacturing cost, agility and most importantly product quality. However, CPM poses new challenges especially on the process monitoring and control aspect of the the manufacturing path \cite{Fonteyne.2015}. One of the most important yet most demanding concepts emergent from CPM research is Quality by Design (QbD), \cite{Yu.2008}. QbD implicated several requirements over all phases ranging from design to monitoring and control of CPM plant. Most importantly the requirements posed on modeling, model-based control, and monitoring of the processes within CPM path. In \cite{ELKHASHAP2020b} the novel QbCon\textsuperscript{\textregistered}1 compact unit for wet granulation and compaction of pharmaceutical mixtures is considered. A Grey box model including a physically motivated structure and data-driven extension is developed, regressed and validated (see also \cite{Meier.2018,Elkhashap2019}). The grey-box model with clear structure comprising a physically-motivated core and a data driven component represents the best possible trade-off between model interpretability/ extrapolability and accuracy/ complexity. However, in order to exploit the proposed model in real-time control further developments are necessary. First, the states of the constructed model are not fully available as measurements at the real plant. Hence an observer design and implementation is necessary. This ensures the availability of real-time estimates for the missing signals which is essential for the feedback model based controller. Moreover as the model comprises a spatially discretized partial differential equation (PDE), the model computational
complexity is significantly high specially for fine spatial discretization. This implicated a high model evaluation complexity and hence a constraint over the computational resources and sampling rate of the employed model-based controller. Model Order Reduction (MOR) of dynamical systems is a very rich field containing numerous methodologies with mainly two principal categories, empirical and system theoretic methods, reader is refereed to \cite{MOROverview1,rafiq2021model} for comprehensive reviews. Methods for nonlinear systems usually are of the first type exploiting analysis techniques on empirical data (often from simulations), e.g. POD Galerkin (\cite{Elkhashap2021}) or Trajectory Piecewise Linear Approximation (TPWL) methods (\cite{MOROverview1}). Such methods, however, suffer from the disadvantage of usually having no guarantees regarding the reduction error, and the preservation of structure/ parametric dependencies and stability properties within the ROMs, see \cite{MOROverview1}. In contrast, system theoretic methods allows for achieving such merits and guarantees even allowing a notion of optimality, e.g \cite{MORbilinH2Zhang.2002}. These methods are however usually applicable to systems of special standard forms as linear or weak nonlinear, see \cite{rafiq2021model,MOROverview1}. 
Hence, the main challenge in order to allow the application of these methods is finding a suitable model formulation allowing its casting into one of these standard forms, e.g. \cite{Elkhashap2022realtime,Elkhashap2021}. Observer design for nonlinear Distributed Parameter Systems (DPS) is still, despite of the numerous efforts, not fully investigated.  The two principle categories for DPS observer design are early lumping and late lumping, see \cite{obs_PDE_Review} for a brief review. In the late lumping approach the observer design procedure utilizes the infinite dimensional (functional space) system directly. First after reaching a formulation of the observer, a finite dimensional approximation (lumping) is performed. Mainly using operator theory, extensions of finite-dimensional observer approaches, e.g. Backstepping (\cite{obs_Backstepping}), sliding mode (\cite{Obs_Slid}), Kalman filtering (\cite{obs_EKF}), are developed. This approach however entails a high level of complexity and is usually constrained to systems of special form (also mainly linear or weak nonlinear). In early lumping approach, the observer design uses a finite dimensional (spatially discrete) approximation of the original system, on which the finite dimensional methods can be directly employed, e.g \cite{obs_PDE_Review}. As the model considered in this contribution is nonlinear and comprises a PDE-DAE cascade, an early lumping approach is pursued. The principal goal of this contribution is twofold: 
\begin{enumerate}
    \item Find a reliable reduced order model (ROM) of the grey-box model proposed in \cite{ELKHASHAP2020b}
    \item Design and evaluate an observer for the estimation of the VFBD unmeasured variables: the spatially distributed moisture content and the dryer mass hold up.
\end{enumerate}
The main methodological novelty in this contribution consists of a special formulation of the solution of the grey-box model PDE part allowing the application of a system theoretic MOR method on the PDE part and finally the synthesis of an observer in the reduced order realm. The model reformulation allows casting the PDE part into a series of a carefully designed vector field and high dimensional Ordinary Differential Equation (ODE) in bilinear input-affine form. The high dimensional bilinear system is then reduced exploiting a $\mathcal{H}_2$ MOR approach preserving the Reduced Order Model (ROM) structure and stability properties also granting the ROM accuracy. Finally, utilizing the constructed ROM an observer design relying on the an Extended Kalman Filter (EKF) approach after proper handling of the algebraic variables and reduction projections is proposed. The paper is organized as follows. First the Grey-box model, its finite-dimensional formulation, and the corresponding MOR approach are introduced. Secondly, the observer design introducing two variants of the observer, one including algebraic state correction, are presented. Third, the evaluation scenarios and configuration are elaborated. Finally, the results are presented followed by a brief conclusion
\section{Methods}
\begin{figure*}[t!]
	\centering
	\def\svgwidth{\textwidth}

	\input{\FigPath Model_Struct.tex}
	\vspace{-0.3cm}
	\caption{diagram of the VFBD under study (left), block diagram illustrating the grey-box model structure (right)} 
	\label{fig:VFBD_sketch_struc}
	\vspace{-0.2cm}
\end{figure*}

\subsection{Grey-box Vibrated Fluid Bed Dryer Model}
The grey-box model representing the full operation of the VFBD is proposed and validated using experimental data in \cite{ELKHASHAP2020b}. 
The model structurally consists of a data-driven part represented by three Gaussian Process Regression (GPR) models mapping the process manipulated variables, i.e. bed vibration intensity $\avib$, drying air mass flow rate $\mdair$, to the coefficients of a conservation Partial Differential Equation (PDE). The VFBD with the unit significant signals as well as a block diagram showing the structure of the grey-box model are depicted in \reffig{fig:VFBD_sketch_struc}. At the core of the grey-box model is the following conservation partial differential equation (PDE) in one spatial dimension $\Svar\in[0,L]$ governing the dry basis moisture content $\cliq(\Svar,t)$ across the dryer bed with length $L$ for any time $t\in \mathbb{R}^{+}$ 

\begin{subequations}\label{eq:PDE}
\begin{equation} \label{eq:PDE_main}
\pdrev{\cliq}{t} = -\vPDE({\GPIn})\pdrev{\cliq}{\Svar} + \DPDE(\GPIn){\pdrev{^2\cliq}{\Svar^2}}-(\phi(\Svar)\vtx{k}{d1} \frac{\mdair}{\mHup}\YDelta +\frac{\dot{\mHup}}{\mHup})\cliq\,,
\end{equation}
\begin{equation} \label{eq:PDE_BC}
\cliq(0,t)=\cliq_{\rm{in}}(t)+\frac{\DPDE}{\vPDE}\pdrev{\cliq}{\Svar}{\bigg|}_{\Svar=0},\quad \pdrev{\cliq}{\Svar}{\bigg|}_{\Svar=L}=0,  
\end{equation}
\end{subequations}
with the granules dry basis moisture content at bed inlet $c_{in}(t)=\frac{\mliq}{\msin}$, the granules effective flow velocity $v$ and diffusion coefficient $D$ as a function of the process manipulated variables $\GPIn=[\mdair,\,\avib]^{\mathsf{T}}$. The spatially dependent function $\phi(z)$ represents the falling drying profile across the bed length with the parameter $\vtx{k}{d1}$ influencing the magnitude of the drying rate (see \cite{ELKHASHAP2020b}). Moreover, the following differential algebraic system representing the mass balance, i.e. solid hold up mass $\mHup$, the bed expansion, as well as the granules drying potential constitutes the remaining physical part of the model   
\begin{equation} \label{eq:ODE}
\drev{\mHup}{t}=\msin-\fTor(\GPIn)\frac{\mHup}{L}\sqrt{2g\,h_b}\,
\end{equation}
\begin{equation} \label{eq:AC1}
g_{1}(h_b,\mHup,\varepsilon,\Delta P,\mdair)=0\,,
\end{equation}
\begin{equation} \label{eq:AC2}
g_{2}(\YDelta,\Tair,\Tsat,\Tdew,\Pair)=0\,,
\end{equation}
with the solid mass flow rate $\msin$ at bed inlet, the bed outlet discharge coefficient $\fTor$, the expanded particle-bed height $h_b$ as a function of the bed porosity $\varepsilon$ and holdup mass $\mHup$, the air pressure loss across the bed height $\Delta P$. The air side drying potential $\Delta Y$ is a function of the inlet air temperature $\Tair$, pressure $\Pair$, saturation temperature $\Tsat$, and relative humidity $\Tdew$. The two algebraic constraints $g_1,\,g_2$ containing the two algebraic variables $\varepsilon,\,\Tsat$ are derived mainly from modified Ergun bed expansion correlation and Mollier h-Y diagram based adiabatic humidification expression (see \cite{ELKHASHAP2020b} for more details). 
The model's data-driven part is constituted mainly from the variables which are intractable to be modelled physically. These are mainly the particles flow parameters. Hence, the PDE coefficients $\vPDE$, $\DPDE$, as well as the bed-outlet discharge factor $\fTor$ are considered as three a-priori uncorrelated maps of the process manipulated variables $\GPIn$ represented by three independent Gaussian Processes $f_{\rm{GP}i} \sim GP(m_i,k_i),\, \forall i \in\{1,2,3\}$ with the corresponding mean $m_i$ and covariance functions $k_i$. Moreover, each of their functional observations are assumed to be affected by additive white noise $\epsilon_i$ 
\begin{equation}\label{eq:GPR}
\vPDE=f_{\rm{GP}1}(\GPIn)+\epsilon_1,\,\DPDE=f_{\rm{GP}2}(\GPIn)+\epsilon_2,\,\fTor=f_{\rm{GP}3}(\GPIn)+\epsilon_3.
\end{equation}
For more information about the GP models construction, regression and validation using experimental data, the reader is referred to \cite{Elkhashap2019,ELKHASHAP2020a}. The model is solved after applying a $\Delta \Svar$ step spatial discretization of the PDE on an $N$ point grid. Defining the state $\vek{x}\in \mathbb{R}^{N+3}$, measured variables $\vek{y}\in \mathbb{R}^2$, manipulated inputs $\vek{u}\in \mathbb{R}^{6}$, and measured disturbances $\vek{w}\in \mathbb{R}^{1}$ vectors
\begin{equation*}
\begin{array}{l}
\vek{x} = [\overbrace{\cliq(0,t), \, \cliq(\Delta \Svar,t), \, \cliq(2\Delta \Svar,t), \, \cdots, \, \cliq(L,t)}^{\let\scriptstyle\textstyle{\vek{x}_1^\mathsf{T}}}, \, \overbrace{\mHup,\, \varepsilon,\Tsat}^{\let\scriptstyle\textstyle{\vek{x}_2^\mathsf{T}}}
]^\mathsf{T},\\ 
\vek{u} = \begin{bmatrix} \Tair, \, \mdair, \, \avib, \, \Delta P, \, \end{bmatrix}^\mathsf{T},\\
\vek{w}= \begin{bmatrix} \msin, \, \mliq,\, \Tdew \end{bmatrix}^\mathsf{T},
\end{array}
\end{equation*} 
the model can be summarized in the following nonlinear descriptor form 
\begin{subequations}\label{eq:DAEsys}
\begin{equation}
\matr{E} \dot{\vek{x}}=\vek{f}(\vek{x},\vek{u},\vek{w})=\begin{pmatrix}\vek{f}_{\rm{PDE}}(\vek{x}_1,\vek{x}_2,\vek{u},\vek{w})\\\vek{f}_{\rm{DAE}}(\vek{x}_2,\vek{u},\vek{w})\end{pmatrix},
\end{equation}
\begin{equation}
\vek{y}=\matr{C}\vek{x}_1,
\end{equation}
\end{subequations}
with the descriptor matrix $\matr{E}=\mathrm{diag}([1,\cdots,1,0,0])$, and measurement matrix $\matr{C}=\mathrm{diag}([0,\cdots,0,1])$. A solver implementation for the numerical integration of the monolithic high dimensional emergent DAE \refeq{eq:DAEsys} using an implicit collocation-based integration scheme is proposed in the original contribution \cite{ELKHASHAP2020a}, where also a brief analysis for the computational cost for different sampling rates is given. Despite the efficient implementation, the computational complexity of the model evaluation imposes a restrain over both the spatial discretization step as well as the time step. This constraint also propagates or is mostly influential in the controller design employing the model. Hence, it was necessary to investigate MOR techniques alleviating the problematic. Hence, reducing the required computational resources and offering more room for different controller design configurations.

\subsection{Model Order Reduction}\label{Sec:MOR}
The main approach adopted for the investigation of potential ROM of the model at hand is based on the structural decomposition of the model and applying system theoretic MOR tools on the parts susceptible to reduction. Examining the model structure, it is obvious that the PDE-part causes the high dimensionality and thus is the most computationaly demanding. This is mainly due to solution distribution along the spatial coordinate and the corresponding spatial discretization. The state space part related to the PDE $\vek{f_{\rm PDE}}$ after applying the spatial discretization reads
\begin{equation}\label{eq:Disc_PDE}
\dot{\vek{x}}_1=(\vPDE\matr{Q}_1+\DPDE\matr{Q}_2+ \vtx{k}{d1}
\frac{\mdair}{\mHup}\Delta Y\matr{Q}_3+\frac{\dot \mHup}{\mHup}\matr{Q}_4) \vek{x}_1 +\vek{b}_1\vPDE\cliq_{\rm{in}}
\end{equation}
the matrices $\matr{Q}_i,\, \forall i\in\{1,\,\cdots,4\}$ emerges from the spatial discretization and finite difference approximation of the partial derivatives 
	\begin{equation}
	\matr{Q}_1=\frac{1}{\Delta \Svar}{\small
		\begin{pmatrix} 
		-1   &0   &\ldots&0\\
		1   & -1  &\ddots&\vdots\\
		\vdots&\ddots&\ddots   &0\\
		0     &\ldots&1   &-1
		\end{pmatrix}},\,
    \matr{Q}_2=\frac{1}{\Delta \Svar^2}{\small
		\begin{pmatrix} 
		-1   &1   &\ldots&0\\
		1   & -2  &\ddots&\vdots\\
		\vdots&\ddots&\ddots   &1\\
		0     &\ldots&1   &-1
		\end{pmatrix}},
	\end{equation}
the matrix $\matr{Q}_3$ can be constructed after sampling the spatial function $\phi(\Svar)$ at the discretization points
\begin{equation}
    \matr{Q}_3=-\mathrm{diag}([\phi(i \Delta \Svar)]),\,\forall i \in \{0,\cdots,N-1\},
\end{equation}
moreover $\matr{Q}_4=-\matr{I}$. The vector $\matr{b}_1=[1,\,0,\cdots]^\mathsf{T}$ is due to the handling of the residual part of the left boundary condition.
Now, defining the following vector field 
\begin{equation}
    \vek{h}(\vek{x}_2,\vek{u},\vek{w})=[\vPDE(\GPIn),\DPDE(\GPIn),\vtx{k}{d1}
\frac{\mdair}{\mHup}\Delta Y,\,\frac{\dot \mHup}{\mHup}-1,\vPDE(\GPIn)\cliq_{\rm{in}}]^\mathsf{T}
\end{equation}
and introducing its output as an augmented input $\Uvec=\vek{h}(\vek{x}_2,\vek{u},\vek{w})$ for the semi-descrite PDE lumps the most significant nonlinearities in the PDE (also the GP model inference) as external inputs. Hence, the semi-discrete PDE \refeq{eq:Disc_PDE} can be cast into a weak nonlinear form, namely a standard bilinear input-affine system
\begin{equation}
    \dot{\vek{x}}_1=\matr{A}\vek{x}_1+\Qf\Uvec\otimes\vek{x}_1+\matr{B}\Uvec,\label{eq:FOM}
\end{equation}
with the bilinear system matrix $\matr{A} \in \mathbb{R}^{\np \times \np}=-\matr{I}$, and input matrix $\matr{B}\in \mathbb{R}^{\np\times 5}=[\matr{0}_{\np\times 4},\,\vek{b}_1]$. The bilinear term is expressed using the Kronecker product notation $\otimes$ with the mode-1 matricization $\matr{\mathcal{Q}}^{[1]}\in \mathbb{R}^{\np\times 5\np}$ of the $3^{rd}$ order tensor $\matr{\mathcal{Q}}\in\mathbb{R}^{\np\times\np\times 5}$ and the frontal slices $\matr{Q}_i \in \mathbb{R}^{\np\times\np}, \forall i\in\{1,..,\udim\}$
\begin{equation*}
\matr{\mathcal{Q}}^{[1]}=[\matr{Q}_1,\,\matr{Q}_2,\,\matr{Q}_3,\,\matr{Q}_4,\,\matr{0}_{\np\times\np}]. 
\end{equation*}
Note that the system \refeq{eq:FOM} is per construction stable\footnote{The system is input to state stable given sufficiently bounded $\ubilin$ and $\matr{Q}_i$. Moreover, the $-1$ eigenvalues with $N$ algebraic multiplicity of the matrix simplifies the stability condition to $\sum\norm{{\matr{Q}_i}}<1/M$ with $M$ being the augmented input vector bound $\norm{\Uvec}<M$} due to the Hurwitz matrix $\vek{A}$. This was an imposed requirement necessary for the further development to the MOR technique employed (see \cite{MORbilinH2Zhang.2002} for further details). Now having achieved a standard bilinear form, the $\mathcal{H}_2$ norm reduction method is directly employed. The method uses a Petrov-Galerkin projection with the trial and test bases $\matr{V},\,\matr{W}\in\mathbb{R}^{\np\times\nr}$ to construct the ROM. Then establishes the first-order necessary conditions of optimality of the $\mathcal{H}_2$ norm of the FOM-ROM error system (see \cite{MORbilinH2Zhang.2002,MORbilinH2} for more details). The generalized Sylvester algorithm proposed in \cite[p.~14]{MORbilinH2} (see \cite{Elkhashap2021} for implementation details) is used with the slight modification in the progression criteria mentioned in \cite{Elkhashap2022model}. After finding the bases corresponding to the minimum $\mathcal{H}_2$ error, the full state $\vek{x}_1$ and reduced state $\xr \in \mathbb{R}^\nr$ vectors are related by the transformation matrices $\matr{V},\matr{T}=(\matr{W}^\mathsf{T}\matr{V})^{-1}\matr{W}^\mathsf{T}$ 
\begin{equation}\label{eq:Proj}
\xr(t)=\matr{T}\vek{x}_1(t) \Leftrightarrow \vek{x}_1(t)=\matr{V}\xr(t),    
\end{equation}
finally the ROM related to the PDE part reads
\begin{subequations}
\begin{equation}
\xdr=\Ar\xr+\Qr \Uvec \otimes \xr + \Br \Uvec,\label{eq:ROM}
\end{equation}	
\begin{equation}
\vek{y}=\Cr \xr
\end{equation}
\end{subequations}
with the reduced state $\xr \in \mathbb{R}^\nr$ of the reduced order $\nr$, system matrices $\Ar \in \mathbb{R}^{\nr\times \nr}$, $\Qr \in \mathbb{R}^{\nr \times \udim\nr}$, $\Br \in \mathbb{R}^{\nr \times \udim}$, and $\Cr \in \mathbb{R}^{1\times \nr}$ 
\begin{align}
&\Ar=\vek{T}\matr{A}\vek{V}=-\matr{I},\quad \Br=\vek{T}\vek{B},\quad\Cr=\vek{C}\vek{V},\\
&\Qr=\Qrf.
\end{align}
Note that the ROM matrix $\Ar$ reduces to be a negative identity matrix this is due to the choice of the augmented vector and the orthonormality of the reduction bases $\matr{W},\matr{V}$. This shows also that the ROM inherits the stability properties of the FOM.
\subsection{Observer Design}\label{Sec:ObsrDes}
The observer design is mainly based on a modified Extended Kalman filter with nonlinear prediction step, where the ROM of the PDE is utilized for a computationally efficient calculation steps in lower dimension. In order to overcome the implications emerging of having a differential system with algebraic constraints, i.e. $\vek{g}=[g_1,\,g_2]^\mathsf{T}$ in (\ref{eq:AC1}-\ref{eq:AC2}), a strategy similar to \cite{EKFDAE1} is adopted. The system model under consideration \refeq{eq:DAEsys} represents an index-1 semi-explicit DAE (Hessenberg index-1 form). Re-partitioning the state vector $\vek{x}$ into a vector for the algebraic variables $\vek{z}=[\epsilon,\, \Tsat]^\mathsf{T}$ and a vector $\xobs$ containing all of the differential variables $\xobs=[\vek{x}_1^\mathsf{T},\, \mHup]^\mathsf{T} $, and re-constructing the vector field in \refeq{eq:DAEsys} accordingly, the following semi-explicit DAE form can be reached
\begin{subequations}\label{eq:DAEsysObs}
\begin{equation} 
\dxobs=\vek{f}_{s}(\xobs,\vek{z},\vek{u},\vek{w})=\begin{pmatrix}\vek{f}_{\rm{PDE}}(\vek{x}_1,\vek{x}_2,\vek{u},\vek{w})\\\vtx{f}{m}(\mHup,\vek{z},\vek{u},\vek{w})\end{pmatrix},
\end{equation}
\begin{equation}
\vek{0}=\vek{g}(\xobs,\vek{z},\vek{u},\vek{w}),
\end{equation}
\end{subequations}
where $\vtx{f}{m}$ represents the right hand side of \refeq{eq:ODE}. The corresponding ROM of the system with the reduced state vector $\xobsr=[\xr^\mathsf{T},\mHup]^\mathsf{T}$ reads
\begin{subequations}\label{eq:DAEsysObs}
\begin{equation} 
\dxobsr=\hat{\vek{f}}(\xobsr,\vek{z},\vek{u},\vek{w})=\begin{pmatrix}\Ar\xr+\Qr \Uvec \otimes \xr + \Br \Uvec,\\\vtx{f}{m}(\mHup,\vek{z},\vek{u},\vek{w})\end{pmatrix},
\end{equation}
\begin{equation}\label{eq:ACsysobs}
\vek{0}=\vek{g}(\xobsr,\vek{z},\vek{u},\vek{w}),
\end{equation}
\end{subequations}
performing a first order taylor-series linearization of the system \refeq{eq:DAEsysObs} produces\footnote{The dependencies on $\vek{u},\,\vek{w}$ are eliminated as they are assumed to be known apriori}
\begin{subequations}
\begin{equation}\label{eq:LinODE}
    \dxobsr=\matr{J}_1(\xobsr,\vek{z})\xobsr+\matr{J}_2(\xobsr,\vek{z})\vek{z}
\end{equation}
\begin{equation}\label{eq:LinAC}
    0=\matr{J}_3(\xobsr,\vek{z})\xobsr+\matr{J}_4(\xobs,\vek{z})\vek{z},
\end{equation}
\end{subequations}
where $\matr{J}_i,\, \forall i \in \{1,\cdots,4\}$ are the corresponding system Jacobians
\begin{equation}\label{eq:JacobiMats}
    \matr{J}_1=\pdrev{\hat{\vek{f}}}{\xobsr},\,\matr{J}_2=\pdrev{\hat{\vek{f}}}{\vek{z}},\, \matr{J}_3=\pdrev{\vek{g}}{\xobsr,},\,\matr{J}_4=\pdrev{\vek{g}}{\vek{z}}, 
\end{equation}
Two variants of the EKF algorithm for the system at hand are considered here (see \cite{EKFDAE1}). The first variant performs an elimination of the algebraic variables, hence, the covariance propagation step for the observer accounts for the differential states only. The mapping matrix $\vtx{\matr{A}}{L1} \in \mathbb{R}^{\nr+1\times \nr+1}$ is calculated by separating $\vek{z}$ in \refeq{eq:LinAC} and substituting the result in \refeq{eq:LinODE}
\begin{equation}
    \dxobsr=\vtx{\matr{A}}{L1}\xobsr,\,
    \vtx{\matr{A}}{L1}=\matr{J}_1 - \matr{J}_2 (\matr{J}_4)^{-1} \matr{J}_3.
\end{equation}
The second variant however considers the algebraic states in the covariance update step. This is done by time-differentiating \refeq{eq:LinAC} once, converting the algebraic equations into differential equations which are then appended to differential system. Hence, the matrix $\vtx{\matr{A}}{L2}\in \mathbb{R}^{\nr+3\times \nr+3}$ represents the linearized system matrix for the augmented vector $\hat{\vec{x}}=[\hat{\vek{\eta}}^\mathsf{T},\,\vek{z}^\mathsf{T}]^\mathsf{T}$ including the algebraic variables 
\begin{equation}
    \dot{\hat{\vec{x}}}=\vtx{\matr{A}}{L2}\hat{\vec{x}},\,
    \vtx{\matr{A}}{L2}=\begin{pmatrix}&\matr{J}_1  &\matr{J}_2\\
                             &-\matr{J}_4^{-1} \matr{J}_3 \matr{J}_1
                              &-\matr{J}_4^{-1} \matr{J}_3 \matr{J}_2
\end{pmatrix},
\end{equation}
The observer design considers the system with discrete measurements (over a uniform sampling period $\Delta t$) affected by white noise $\nu_k$
\begin{equation}
    \vek{y}_{k+1}=\Cobs\hat{\vek{x}}+\nu_k,\, \Cobs=[\Cr,\vek{0}_{1\times3}],
\end{equation}
The covariance update step is then preformed in discrete time exploiting the matrix exponential for the calculation of the corresponding transition matrix 
\begin{equation}\label{eq:TransMat}
    \matr{\Phi}_1=\mathrm{exp}(\vtx{\matr{A}}{L1} \Delta t),\, \matr{\Phi}_2=\mathrm{exp}(\vtx{\matr{A}}{L2} \Delta t)
\end{equation}
Hereafter, the algorithm for the second case will be only elaborated being the one requiring more treatment. The first case can then be naturally deduced following the same steps omitting the algebraic states from the covariance update and correction steps. In order to run the EKF steps in the reduced order dimension, mappings for the states and covariance matrix are needed to project the states first to the reduced order. After executing the observer steps, the full system states and covariance matrix are recovered using the inverse projection. The state projections are fully defined through \refeq{eq:Proj} which are utilized in the construction of the observer mappings. These require the projection of certain blocks related to the states concerned with the reduction. The mappings can be represented fully using the projection matrices $\vek{\Gamma},\vec{\Gamma}^{-1}$
\begin{equation}
    \matr{\Gamma}=\begin{pmatrix}
      \matr{T} & \bigzero_{\nr\times3}\\
     \bigzero_{3\times\np} & \Imat{3}\\
    \end{pmatrix},\,
    \matr{\Gamma}^{-1}=\begin{pmatrix}
      \matr{V} & \bigzero_{\np\times3}\\
     \bigzero_{3\times\nr} & \Imat{3}\\
    \end{pmatrix}
\end{equation}
Moreover, considering state white noise vector $\vek{\omega}_k \in \mathbb{R}^{\np+1}$ with no consideration for separate independent noise terms for the algebraic variables, the noise matrix $\matr{\Omega}_k\in \mathbb{R}^{{\nr+3}\times{\nr+3}}$ is calculated in the reduced order dimension
\begin{equation}\label{eq:NoisMat}
    \matr{\Omega}_k=\matr{\Psi}
    \begin{pmatrix}
      \matr{T} & \vec{0}\\
      \vec{0}^\mathsf{T} & 1\\
    \end{pmatrix}
    \mathrm{diag}(\vek{\omega}_k)
    \begin{pmatrix}
      \matr{V} & \vec{0}\\
      \vec{0}^\mathsf{T} & 1\\
    \end{pmatrix}
    \matr{\Psi}^\mathsf{T},\, 
    \matr{\Psi}=    \begin{pmatrix}
      \Imat{\nr+1}\\
      \matr{J}_4^{-1}\matr{J}_3\\
    \end{pmatrix}
\end{equation}
Now, at a certain time index $k$ given the measurement of the moisture content at the bed outlet $y^*_k$, previous full order state estimate $\hat{\vek{x}}_{k-1}$ and covariance matrix $\Pmat_{k-1}$, current sample inputs $\vek{u}_{k}$, measured disturbances $\vek{w}_{k}$, state noise $\vek{\omega}_k$, and measurement noise $\nu_k$, the observer steps can be summarized as follows:
\begin{algorithm}
\caption{ROM EKF with nonlinear prediction}\label{alg:cap}
\hspace*{\algorithmicindent} \textbf{Input}: $\matr{P}_{k-1},\hat{\vek{x}}_{k-1},\vek{u}_{k},\vek{w}_{k},y^*_k,\vek{\omega}_{k},\nu_{k} $\\
\hspace*{\algorithmicindent} \textbf{Output}:$\matr{P}_{k},\vek{x}_k$ 
\begin{algorithmic}[1]
\State  $\hat{\vek{x}}_{k-1}=\matr{\Gamma}\vek{x}_{k-1}$ \Comment{Project FOM states}
\State $\hat{\Pmat}_{k-1}=\matr{\Gamma}\Pmat_{k-1}\matr{\Gamma}^{-1}$ \Comment{Project FOM covariance}

\State $\hat{\vek{x}}_k=$ integration step of \refeq{eq:DAEsysObs} \Comment{nonlinear prediction}
\State Evaluate $\matr{J}_{1,...,4}$  at $\hat{\vek{x}}_k,\vek{u}_{k},\vek{w}_{k}$
\State $\matr{\Phi}_2=\mathrm{exp}(\vtx{\matr{A}}{L2}(\hat{\vek{x}}_k,\vek{u}_{k},\vek{w}_{k})\Delta t)$\Comment{transition matrix}
\State Calculate $\matr{\Omega_k}$ according to \refeq{eq:NoisMat} \Comment{noise matrix}
\State $\hat{\Pmat}_{k}=\matr{\Phi}_2\hat{\Pmat}_{k-1}\matr{\Phi}_2^\mathsf{T}+\matr{\Omega}_k$ \Comment{Covariance Propagation}

\State $\matr{K} = \hat{\matr{P}}_k\Cobs(\Cobs \hat{\matr{P}}_k^-\Cobs^\mathsf{T} + \nu_k)^{-1}$ \Comment{Kalman Gain}
\State $\hat{\vek{x}}_k\gets\hat{\vek{x}}_k + K(y^* - \Cobs\hat{\vek{x}}_k) $ \Comment{State Correction}
\State $\hat{\matr{P}}_k\gets(\matr{I}-\matr{K}\Cobs)\hat{\matr{P}}_k$\Comment{Covariance Correction} 

\State update $\vek{z}_k$ by resolving \refeq{eq:ACsysobs} with initial value $\hat{\vek{x}}_k$ using fast Newton \Comment{Reconcile Algebraic States} 
\State  $\vek{x}_{k}=\matr{\Gamma}^{-1}\hat{\vek{x}}_{k}$\Comment{Project ROM states}
\State  $\Pmat_{k}=\matr{\Gamma}^{-1}\hat{\Pmat}_{k}\matr{\Gamma}$\Comment{Project ROM covariance}
\end{algorithmic}
\end{algorithm}

\section{Experiments}
In order to evaluate the methods presented in the previous section, the model experimental data used for validation in \cite{ELKHASHAP2020b} as well as certain designed test scenarios are utilized. The experimental data used comprises $3\,\mathrm{hr}$ operation of VFBD including a wide range of variations in the process conditions (inputs/ measured disturbances and states). The two main points considered for evaluation are:
\begin{itemize}
    \item Comparison of the ROM constructed using the approach elaborated in \refsec{Sec:MOR} against the FOM
    \item Evaluation of the two variants of the observer design proposed in \refsec{Sec:ObsrDes} also against the FOM-based observer
\end{itemize}
The cpu time is recorded for each step. All of the simulations required for the evaluations are performed on a Windows 10 PC (Intel(R) Core(TM) i7-7700HQ CPU@2.8GHZ, 8GB RAM). For the numerical integration of the DAEs a highly efficient implementation of an implicit numerical integration scheme using casADi (\cite{Andersson2019}). Namely a collocation scheme with 3rd order Legendre polynomials using 3 collocation points is used. This ensures the stability of the numerical integration even for arbitrary time step $\Delta t$ as the scheme is $A$-stable. Moreover,  casADi algorithmic differentiation is exploited for efficient representation and evaluation of the system jacobians. Finally to increase the efficiency a C-code generation of the implementations is utilized for all of the modules presented.
\subsection{ROM Evaluation}
In the first evaluations the ROM constructed through the approach proposed in \refsec{Sec:MOR} is evaluated against the FOM in simulations. Both of the models are constructed using the previously mentioned numerical scheme with $\Delta t=2\,\mathrm{sec}$ matching the experimental data sampling rate. The FOM is generated using $N=1000$ representing a very fine resolution across the be length ($\Delta \Svar= \frac{L}{1000}$), this is to demonstrate the potential of the method specially for up-scaled applications. The ROM dimension $r=7$ is shown to be sufficient for the accurate reconstruction of the system dynamics (empirically through simulations). 
\subsection{Observer Evaluation}
For both of the observer variants further model related aspects are handled in the implementation. For example, physical plausibility of the states and algebraic variables are ensured after each observer step by sanity checks and projections of the variables to their physically-permissible spaces, e.g. $\mHup \in \mathbb{R}^{+}, \varepsilon \in [0,1]$. Also checks for numerical faults, division by zeros are considered and safe-guarded numerically in implementation. In order to evaluate the observer design approach elaborated in \refsec{Sec:ObsrDes}, three evaluation scenarios are carried out:
\begin{enumerate}
    \item Compare the two variants of proposed observer evaluating their accuracy, robustness against parameterization, e.g. initialization, and computational efficiency
    \item Evaluate the second variant against an equivalent version utilizing the FOM (quantify the reduction effect) 
    \item Evaluate the performance of the second variant on experimental data of the real operation of the unit.
\end{enumerate}
 For the first case, the same 3 hr experimental data is used with the observer design variant where the algebraic variables correction step is considered.  
 The ground truth of the states and the algebraic variables are calculated using forward simulation of the full order validated model \cite{ELKHASHAP2020b}. The measurement noise variance is chosen according to the empirically determined maximum standard deviation of the redundant loss on drying measurement, i.e. $\nu=0.006^2$ (see \cite{Elkhashap2019}). In the second and third case, the candidates of the proposed observers are compared regarding their computation time, accuracy (deviation from ground truth), convergence speed, and sensitivity to parameterization. The observer parameters, i.e. initial covariance matrix, are fixed for the candidates but chosen randomly for the 100, 600 runs each with the duration of 2, 1 minutes investigating the transient period of the observer. A random state and algebraic variables initialization for all runs are also chosen in order to investigate the sensitivity of the observer variant to initial guesses. 
\section{Results}
\begin{figure}[h!]

  \centering
    \includegraphics[width=\linewidth]{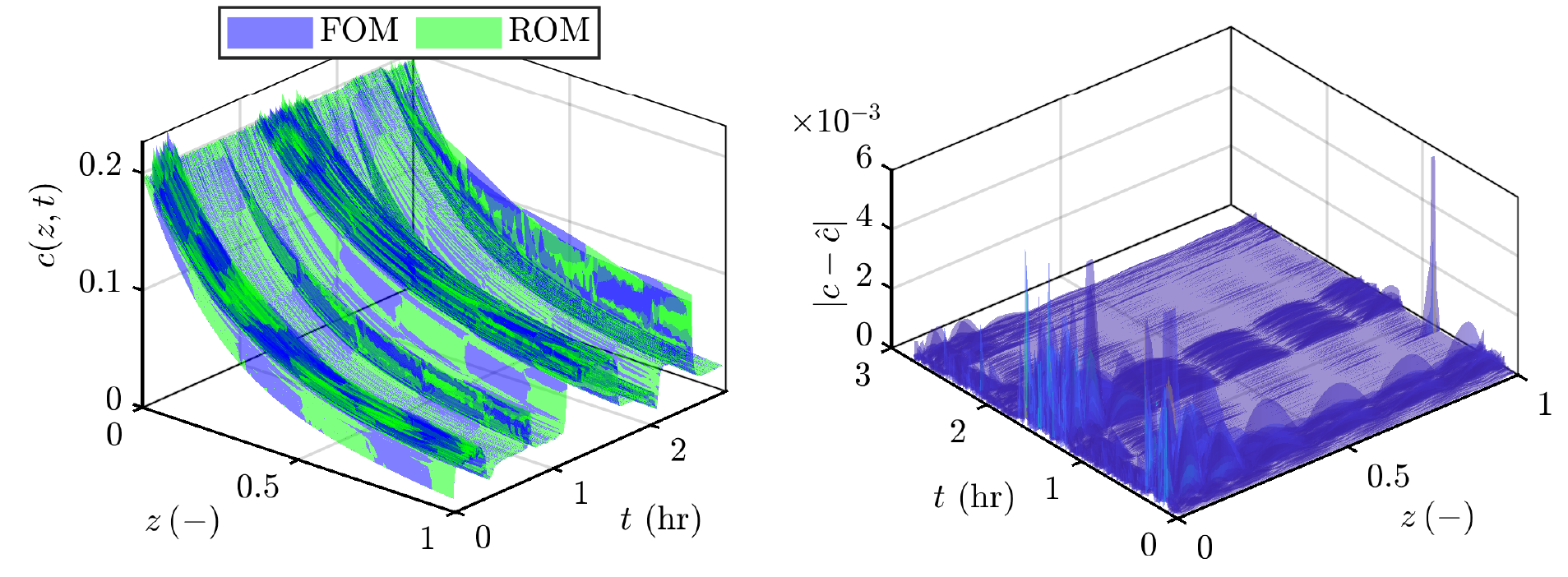}
    \vspace{-0.5cm}
  	\caption{FOM vs ROM moisture content prediction (left) and the corresponding spatio-temporal absolute error (right) for the validation experiments}
	\vspace{-0.2cm}
	\label{fig:FOMvsROMValid}
\end{figure}

In \reffig{fig:FOMvsROMValid} a 3d visualization of the moisture content across the bed length for the 3 hrs of the experiment is shown. The ROM generated using the approach elaborated above could reconstruct the full order state of the system efficiently with high accuracy. Moreover the absolute point-wise error of the ROM moisture content prediction is shown in the right plot. It can be observed that the absolute moisture content error of the ROM across the bed length and the 3 hr experiment did not exceed $5\times10^{-3}\, \mathrm{(kg/kg)}$. This shows that the ROM accuracy is high enough with practically irrelevant reduction error. The cpu time needed by both FOM and ROM for one-step prediction for different $\np$ is shown in \reftab{tab:Results}. For the $\np=1000$ resolution ROM the mean cpu time for one step computation is $1.1\, \mathrm{msec}$ in comparison to a $39.5 \, \mathrm{msec}$ achieving a speedup of almost a factor of 40. 

\begin{figure}[h!]
  \centering
    \includegraphics[width=\linewidth]{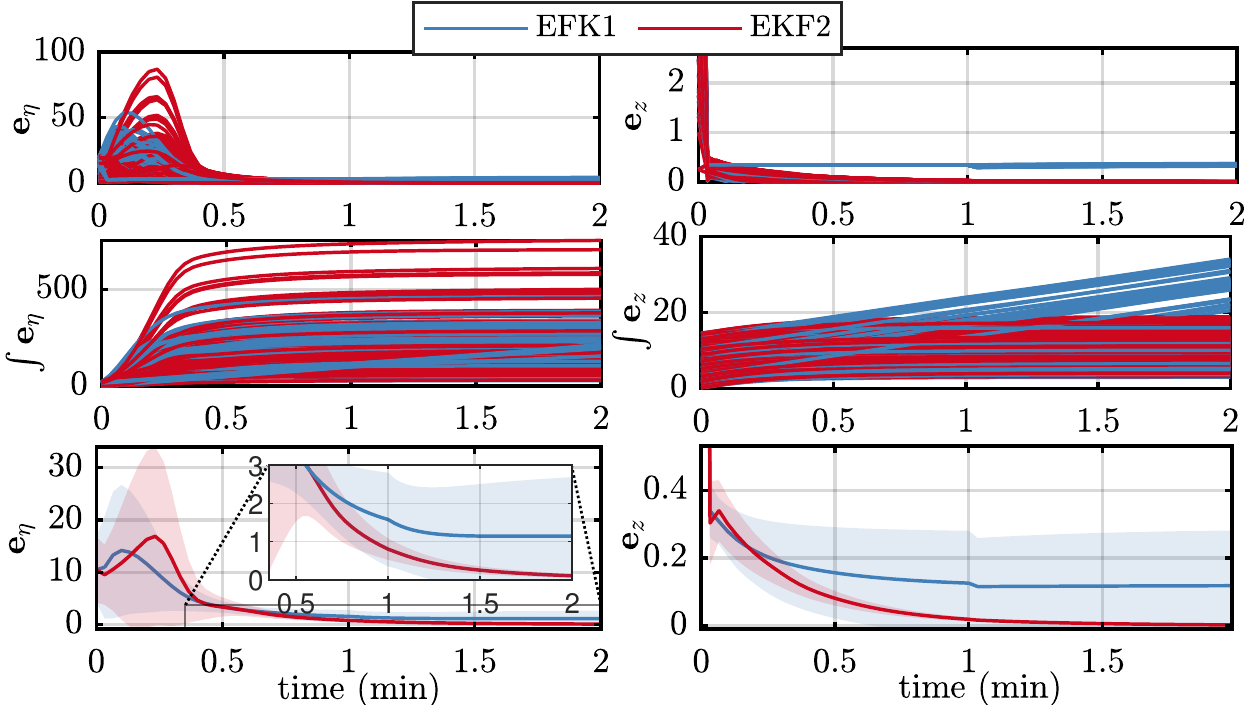}
    \vspace{-0.7cm}
  	\caption{ROM observers state estimation errors (top left) and algebraic variables errors (top right) for the 100 random runs for observer variant 1 (blue) and 2 (red); Errors time integral (middle plots); 100 runs errors mean as solid line and standard deviation as shaded area (bottom plots)}
	\vspace{-0.2cm}
	\label{fig:EKF1}
\end{figure}

\begin{figure}[h!]
  \centering
    \includegraphics[width=\linewidth]{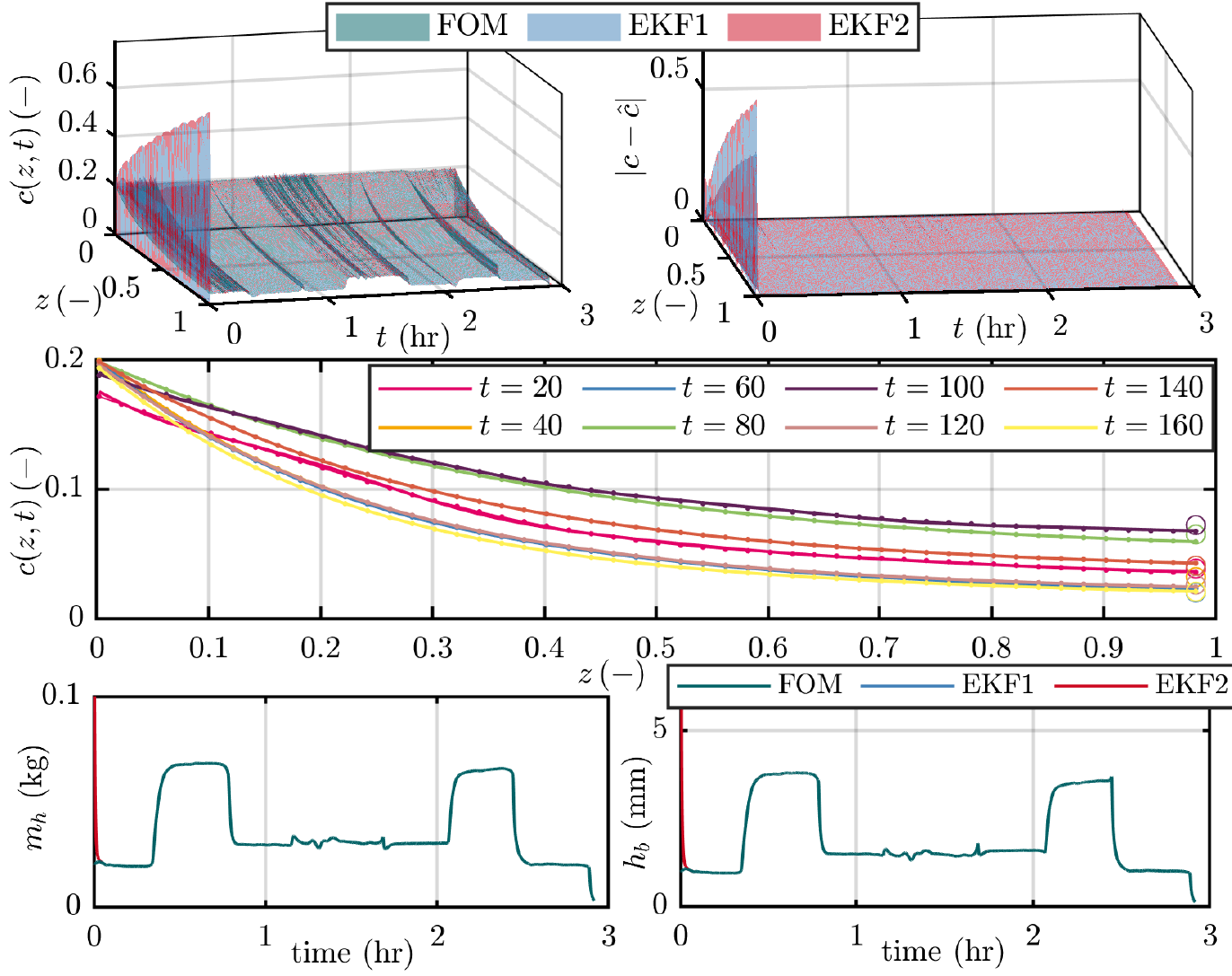}
    \vspace{-0.7cm}
  	\caption{Observer moisture content estimation results (top left) and the corresponding spatio-temporal errors (top right); FOM (solid line), observer variant 1 (dashed), and 2 (dotted) moisture content spatial profiles at different time points (middle), moisture measurements available to the observers are visualized with o-markers; mass holdup (bottom left) and expanded bed height (bottom right) estimated vs true values.}
	\vspace{-0.1cm}
	\label{fig:EKF2}
\end{figure}

\begin{figure}[h!]
  \centering
    \includegraphics[width=\linewidth]{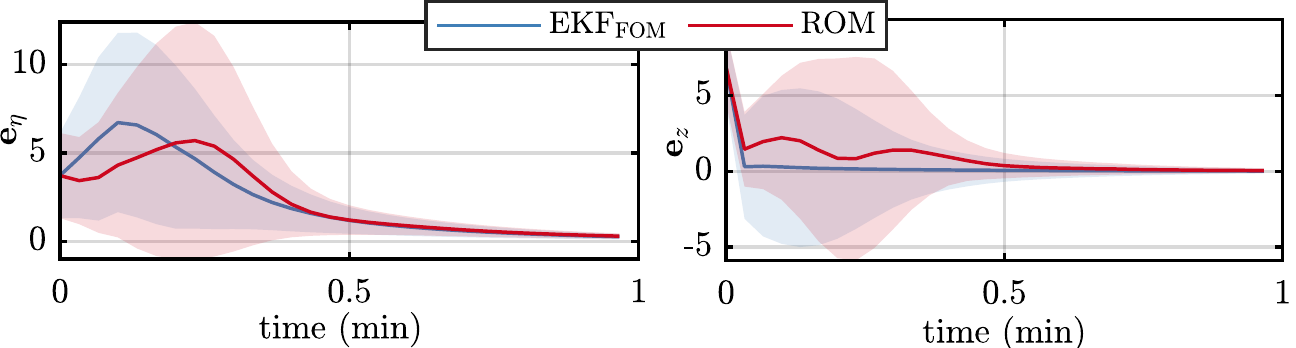}
    \vspace{-0.8cm}
  	\caption{FOM based observer (blue) and ROM observer (red) states (left) and algebraic variables (right) estimation error for all random runs; mean as solid line and standard deviation as shaded area}
	\label{fig:EKF_FOMvsROM}
\end{figure}

\reffig{fig:EKF1} depicts the results of the 100 observer random runs comparing the two ROM-based observer variants. It can be observed that the second variant considering the algebraic variables in update step is of superior performance. This is mainly as its error converged for all of the runs to the correct states and algebraic variables values (see middle plot). On the other hand the first variant could not converge to the true algebraic variables for some cases causing a steady state error in state estimation (see bottom plots in \reffig{fig:EKF1}). This indicates that a correction mechanism of the algebraic variables is necessary. The mean computation time of a single step for the first variant is $15.4\, \mathrm{msec}$ in comparison to $17.5\, \mathrm{msec}$ of the second variant both with standard deviation below $1.9\,\mathrm{msec}$ (for $\np=500$). This shows that the benefit achieved from the including the algebraic variables correction comes with a slight increase in the computational cost. The performance of the two observer variants on the experimental data is shown in \reffig{fig:EKF2}. It can be observed that both variants converged to the true moisture content in the first minutes of the experiments with practically negligible error.    
Moreover, \reffig{fig:EKF_FOMvsROM} illustrates the state and algebraic variables estimation error of the second variant ROM-based observer and its FOM-based equivalent for the 600 random runs. 
As expected, the convergence of the estimation error is faster for the FOM-based observer as it comprises no reduction inaccuracy. However, the difference in performance is minimal and most importantly no fail-runs were detected (observer converged to true value for all 600 runs). \reftab{tab:Results} summarizes the cpu-time needed for the single step evaluation of the various observers. The FOM-based observer one-step time for $N=100$ reached $18.8 \,\mathrm{sec}$ compared to $4.6 \,\mathrm{msec}$ of the ROM-based observer. This indicates a massive computational reduction of more that three orders of magnitude ($4.7 \times 10^3$) to the FOM-based observer.
\newcolumntype{L}{>{$}l<{$}}
\newcolumntype{C}{>{$}c<{$}}
\newcolumntype{R}{>{$}r<{$}}
\newcommand{\vln}[1]{\multicolumn{1}{C}{#1}}
\begin{table}[h!]
	\caption{one step computation time mean and standard deviation in $\mathrm{msec}$ of the FOM, ROM, and observers for $\nr=7$ and different $\np$}
	\vspace{-0.1cm}
	\setlength{\tabcolsep}{2.5pt} 
	\renewcommand{\arraystretch}{0.6} 
	\centering
	\begin{tabular}{LCCCCCCCCCC}
		\toprule
		\multicolumn{1}{C}{\multirow{2}{*}{$N$}} & \multicolumn{2}{C}{{\mathrm{FOM} }} & \multicolumn{2}{C}{{\mathrm{ROM} }} &
		\multicolumn{2}{C}{{\mathrm{EKF}_\mathrm{FOM}}}&
		\multicolumn{2}{C}{{\mathrm{EKF1}}}&
		\multicolumn{2}{C}{{\mathrm{EKF2}}}\\ \cmidrule(r){2-3} \cmidrule(r){4-5} \cmidrule(r){6-7}  \cmidrule(r){8-9} \cmidrule(r){10-11}  
		& \mathrm{mean}  & \mathrm{std}  & \mathrm{mean}  & \mathrm{std}& \mathrm{mean}  & \mathrm{std}& \mathrm{mean}  & \mathrm{std}& \mathrm{mean}  & \mathrm{std}\\ \cmidrule(r){1-11}

         \vln{ 20 }  & 1.20 & 0.14 & 1.02 & 0.17 & 48.06& 3.64& 2.30& 0.64& 4.53 & 0.56\\ 
         \vln{ 30 }  & 1.42 & 0.09 & 0.96 & 0.06 & 192.1& 15.8& 2.57& 0.29& 4.20 & 0.40\\ 
         \vln{ 50 }  & 2.12 & 0.24 & 0.99 & 0.10 & 1510 & 168 & 7.54& 0.90& 14.0 & 1.98\\ 
         \vln{ 60 }  & 2.23 & 0.14 & 1.03 & 1.51 & 2870 & 230 & 2.06& 0.78& 4.39 & 0.42\\ 
         \vln{ 90 }  & 3.06 & 0.17 & 0.94 & 0.06 & 12519& 1146& 2.29& 0.81& 4.51 & 0.43\\ 
         \vln{ 100 } & 3.47 & 0.16 & 0.94 & 0.06 & \Hi18876& 1516& 3.07& 0.60& \Hi4.63 & 0.43\\ 
         \vln{ 200 } & 6.18 & 0.65 & 0.96 & 0.13 &    - &   - & 4.54& 0.66& 6.16 & 0.60\\ 
         \vln{ 500 } & 14.7 & 1.10 & 0.97 & 0.08 &    - &   - & \Hi15.4& 1.84& \Hi17.5 & 1.35\\ 
         \vln{ 10^3 }& \Hi39.9 & 5.59 & \Hi1.1  & 0.12 &    - &   - & 67.1& 11.2& \Hi72.1 & 6.77\\

		\bottomrule
	\end{tabular}\label{tab:Results}
\vspace{-0.1cm}
\end{table}
This most demanding computational step is the symbolic evaluation of the matrix exponential, which is also intractable for a FOM with $N>200$ as the required memory needed for code generation explodes. Hence, it can be concluded that the ROM-observer proposed represents a pivotal leap producing a comparable performance of the FOM-based observer with a significantly reduced computational cost.       

\section{Conclusion}
ROM-based observers pose a relevant method for PDE state observation even when coupled with differential algebraic systems. A novel Model Order Reduction (MOR) method for the grey-box model of the vibrated fluid bed dryer (VFBD) is proposed. The MOR strategy is based on a special decomposition of the model designing a vector field mapping the coupling between the rest of the model and the PDE part. This vector field along with a special formulation of the PDE spatial-discretization allows casting the PDE part into a bilinear input affine form which is then reduced using an $\mathcal{H}_2$ norm method. Based on the system ROMs an observer design for the model states including the spatially distributed moisture content is proposed. The observer design relies on an Extended Kalman Filtering strategy with nonlinear prediction in the reduced order state. Two variants of the observer are proposed, in order to handle also the algebraic variables and include a corresponding correction mechanism. Both the reduction and observer approaches are evaluated using experimental data and Monte-Carlo simulations. Results show, that the proposed algorithms, i.e. observer, prediction models, can be executed in a real-time with sampling periods in milliseconds. Applying the developed observers allows a distributed state observation for the complex VFBD process along with the differential and algebraic states with practically negligible error.


\bibliography{ifacconf}             

\begin{thebibliography}{20}
\providecommand{\natexlab}[1]{#1}
\providecommand{\url}[1]{\texttt{#1}}
\providecommand{\urlprefix}{URL }
\expandafter\ifx\csname urlstyle\endcsname\relax
  \providecommand{\doi}[1]{doi:\discretionary{}{}{}#1}\else
  \providecommand{\doi}{doi:\discretionary{}{}{}\begingroup
  \urlstyle{rm}\Url}\fi

\bibitem[{Afshar et~al.(2020)Afshar, Germ, and Morris}]{obs_EKF}
Afshar, S., Germ, F., and Morris, K. (2020).
\newblock Well-posedness of extended kalman filter equations for semilinear
  infinite-dimensional systems.
\newblock In \emph{2020 59th IEEE Conference on Decision and Control (CDC)},
  1210--1215.

\bibitem[{Andersson et~al.(2019)Andersson, Gillis, Horn, Rawlings, and
  Diehl}]{Andersson2019}
Andersson, J.A.E., Gillis, J., Horn, G., Rawlings, J.B., and Diehl, M. (2019).
\newblock {CasADi} -- {A} software framework for nonlinear optimization and
  optimal control.
\newblock \emph{Mathematical Programming Computation}, 11(1), 1--36.

\bibitem[{Benner and Breiten(2012)}]{MORbilinH2}
Benner, P. and Breiten, T. (2012).
\newblock Interpolation-based $\mathcal{H}_2$-model reduction of bilinear
  control systems.
\newblock \emph{SIAM Journal on Matrix Analysis and Applications}, 33(3),
  859--885.

\bibitem[{Elkhashap et~al.(2020{\natexlab{a}})Elkhashap, Meier, Stenger, and
  Abel}]{ELKHASHAP2020a}
Elkhashap, A., Meier, R., Stenger, D., and Abel, D. (2020{\natexlab{a}}).
\newblock Grey-box approach for the prediction of variable residence time
  distribution in continuous pharmaceutical manufacturing.
\newblock \emph{IFAC-PapersOnLine}, 53(2), 10360--10365.
\newblock 21st IFAC World Congress.

\bibitem[{Elkhashap and Abel(2021)}]{Elkhashap2021}
Elkhashap, A. and Abel, D. (2021).
\newblock Parametric model order reduction of variable parameter axial
  dispersion model.
\newblock In \emph{2021 IEEE Conference on Control Technology and Applications
  (CCTA)}, 408--415.

\bibitem[{Elkhashap and Abel(2022)}]{Elkhashap2022model}
Elkhashap, A. and Abel, D. (2022).
\newblock Model order reduction of advection-dispersion-reaction equation with
  time-varying coefficients, application to real-time water quality monitoring.
\newblock In \emph{2022 European Control Conference (ECC)}.
\newblock (accepted).

\bibitem[{Elkhashap et~al.(2019)Elkhashap, Meier, and Abel}]{Elkhashap2019}
Elkhashap, A., Meier, R., and Abel, D. (2019).
\newblock {M}odeling and {C}ontrol of a {C}ontinuous {V}ibrated {F}luidized
  {B}ed {D}ryer in {P}harmaceutical {T}ablets {P}roduction.
\newblock \emph{Die pharmazeutische Industrie : pharmind}, 81(12), 1693--1700.
\newblock This article was first published in VDI-Berichte Nr. 2351,
  VDI-Kongress Automation 2019, Düsseldorf: VDI 2019.

\bibitem[{Elkhashap et~al.(2020{\natexlab{b}})Elkhashap, Meier, and
  Abel}]{ELKHASHAP2020b}
Elkhashap, A., Meier, R., and Abel, D. (2020{\natexlab{b}}).
\newblock A grey box distributed parameter model for a continuous vibrated
  fluidized bed dryer in pharmaceutical manufacturing.
\newblock In \emph{2020 European Control Conference (ECC)}, 1415--1421.

\bibitem[{Elkhashap et~al.(2022)Elkhashap, Rüschen, and
  Abel}]{Elkhashap2022realtime}
Elkhashap, A., Rüschen, D., and Abel, D. (2022).
\newblock Real-time monitoring and control of water networks.
\newblock In \emph{2022 American Control Conference (ACC)}.
\newblock (accepted).

\bibitem[{Fonteyne et~al.(2015)Fonteyne, Vercruysse, de~Leersnyder, {van
  Snick}, Vervaet, Remon, and de~Beer}]{Fonteyne.2015}
Fonteyne, M., Vercruysse, J., de~Leersnyder, F., {van Snick}, B., Vervaet, C.,
  Remon, J.P., and de~Beer, T. (2015).
\newblock Process analytical technology for continuous manufacturing of
  solid-dosage forms.
\newblock \emph{TrAC Trends in Analytical Chemistry}, 67, 159--166.

\bibitem[{Krstic and Smyshlyaev(2008)}]{obs_Backstepping}
Krstic, M. and Smyshlyaev, A. (2008).
\newblock \emph{Boundary control of PDEs: A course on backstepping designs}.
\newblock SIAM.

\bibitem[{Mandela et~al.(2009)Mandela, Rengaswamy, and Narasimhan}]{EKFDAE1}
Mandela, R.K., Rengaswamy, R., and Narasimhan, S. (2009).
\newblock Nonlinear state estimation of differential algebraic systems.
\newblock \emph{IFAC Proceedings Volumes}, 42(11), 792--797.
\newblock 7th IFAC Symposium on Advanced Control of Chemical Processes.

\bibitem[{Meier and Emanuele(2018)}]{Meier.2018}
Meier, R. and Emanuele, D. (2018).
\newblock Kontinuierliche feuchtgranulierung und wirbelschichttrocknung:
  Experimentelle untersuchung eines neuartigen, revolutionaeren systems
  prozesstechnik.
\newblock \emph{TechnoPharm}, 8(3), 124.
\newblock In german.

\bibitem[{Miranda et~al.(2012)Miranda, Moreno, Chairez, and Fridman}]{Obs_Slid}
Miranda, R., Moreno, J.A., Chairez, J., and Fridman, L. (2012).
\newblock Observer design for a class of hyperbolic pde equation based on a
  distributed super twisting algorithm.
\newblock In \emph{2012 12th International Workshop on Variable Structure
  Systems}, 367--372.

\bibitem[{Rafiq and Bazaz(2021)}]{rafiq2021model}
Rafiq, D. and Bazaz, M.A. (2021).
\newblock Model order reduction via moment-matching: a state of the art review.
\newblock \emph{Archives of Computational Methods in Engineering}, 1--21.

\bibitem[{Rantanen and Khinast(2015)}]{Rantanen.2015}
Rantanen, J. and Khinast, J. (2015).
\newblock The future of pharmaceutical manufacturing sciences.
\newblock \emph{Journal of pharmaceutical sciences}, 104(11), 3612--3638.

\bibitem[{{Ulrike Baur} et~al.(2014){Ulrike Baur}, {Peter Benner}, and {Lihong
  Feng}}]{MOROverview1}
{Ulrike Baur}, {Peter Benner}, and {Lihong Feng} (2014).
\newblock Model order reduction for linear and nonlinear systems: A
  system-theoretic perspective.
\newblock \emph{Archives of Computational Methods in Engineering}, 21(4),
  331--358.

\bibitem[{Yu(2008)}]{Yu.2008}
Yu, L.X. (2008).
\newblock Pharmaceutical quality by design: product and process development,
  understanding, and control.
\newblock \emph{Pharmaceutical research}, 25(4), 781--791.

\bibitem[{Yupanqui~Tello et~al.(2021)Yupanqui~Tello, Vande~Wouwer, and
  Coutinho}]{obs_PDE_Review}
Yupanqui~Tello, I.F., Vande~Wouwer, A., and Coutinho, D. (2021).
\newblock A concise review of state estimation techniques for partial
  differential equation systems.
\newblock \emph{Mathematics}, 9(24), 3180.

\bibitem[{Zhang and Lam(2002)}]{MORbilinH2Zhang.2002}
Zhang, L. and Lam, J. (2002).
\newblock On h2 model reduction of bilinear systems.
\newblock \emph{Automatica}, 38(2), 205--216.

\end{thebibliography}




\end{document}